\documentstyle[preprint,aps]{revtex}

\begin{document}
\draft
\title{Enhanced Transmission Through Disordered Potential Barrier}
\author{V. Freilikher, M. Pustilnik and I. Yurkevich}
\address{The Jack and Perl Resnick Institute of Advanced Technology,\\
Department of Physics, Bar-Ilan University, Ramat-Gan 52900, Israel}
\date{\today }
\maketitle

\begin{abstract}
Effect of weak disorder on tunneling through a potential barrier is studied
analytically. A diagrammatic approach based on the specific behavior of
subbarrier wave functions is developed. The problem is shown to be
equivalent to that of tunneling through rectangular barriers with Gaussian
distributed heights. The distribution function for the transmission
coefficient $T$ is derived, and statistical moments $\left\langle
T^n\right\rangle $ are calculated. The surprising result is that in average
disorder increases both tunneling conductance and resistance.
\end{abstract}


\narrowtext

It is well known that in one-dimensional disordered system the interference
of propagating waves results in strong localization. Transmission
coefficient (transmittance) $T$ for such a system is not self-averaging
quantity and obeys the log-normal distribution law.

The question arises what is the effect of multiply scattering of subbarrier
(evanescent) waves on the tunneling transmissivity, when the standard
conceptions of phases, interference and coherent trajectories are
unapplicable. This problem apparently arises in a wide variety of
applications in solid state theory, optics and radiophysics.

The particular question this paper is addressed to is how the
one-dimensional quantum tunneling is disturbed by random perturbations of
the shape of potential barrier. We consider a disordered region $0\leq
x\leq L$ with the potential given by
\begin{equation}
U(x)=V_0+v(x)\text{ ,}
\end{equation}
$v(x)$ being a random function of $x$. An incident (from the left) particle
has energy $k^2$ (we use units where $\hbar ^2/2m=1$) that is less then the
height of the unperturbed barrier $V_0$:
\begin{equation}
V_0=k^2+\kappa ^2\;>k^2.
\end{equation}
Transmittance of such a system is expressed through the Green function of
the problem as
\begin{equation}
T=|\tau |^2=4k^2\left| {\cal G}(0,L)\right| ^2,
\end{equation}
where $\tau $ is the amplitude transmission coefficient, and ${\cal G}%
(x,x^{\prime })$ satisfies the equation
\begin{equation}
\left( k^2-H\right) {\cal G}(x,x^{\prime })=\delta (x-x^{\prime }),
\label{eq.11}
\end{equation}
with the Hamiltonian
\begin{equation}
H=-\frac{d^2}{dx^2}+U(x)\left[ \theta (x)-\theta (x-L)\right] .
\label{eq.11a}
\end{equation}

In the case $V_0<k^2$ there exist standard approaches based on the averaging
over rapidly oscillating phases \cite{classics}. In the case of subbarrier
penetration evanescent waves have no rapidly oscillating phases, and we need
another method allowing to utilize their special features. Below we develope
a diagrammatic perturbation approach that takes an advantage of the rapid
decay of evanescent waves.

In order to take into account explicitly scattering by the barrier edges we
decompose $H$ into 3 terms so that
\begin{equation}
H=\left( -\frac{d^2}{dx^2}+V_0\right) +\Delta V_1(x)+\Delta V_2(x)\;,
\end{equation}
where
\[
\Delta V_1(x)=v(x)\left[ \theta (x)-\theta (x-L)\right] \;,
\]
\[
\Delta V_2(x)=V_0\left[ \theta (x)-\theta (x-L)-1\right] \;.
\]
We introduce the ''subbarrier'' Green function $G(x,x^{\prime })$ satisfying
the equation
\begin{equation}
\left( \frac{\partial ^2}{\partial x^2}-\kappa ^2-\Delta V_1(x)\right)
G(x,x^{\prime })=\delta (x-x^{\prime }).\label{eq.6}
\end{equation}
Then the scattering by edges described by the term $\Delta V_2(x)$ can be
taken into account through the equation
\begin{equation}
{\cal G}(x,x^{\prime })=G(x,x^{\prime })+\int_{-\infty }^\infty
dx_1G(x,x_1)\Delta V_2(x_1){\cal G}(x_1,x^{\prime })\;.\label{eq.7}
\end{equation}
Since $\Delta V_2(x_1)=0$ for $0\leq x_1\leq L$ no scattering take place in
the region of integration in Eq.(\ref{eq.7}), and the behavior of both $%
G(x,x_1)$ and ${\cal G}(x_1,x^{\prime })$ is completely defined by their
values at the edges. For $x,x^{\prime }\in (0,L)$ we obtain from Eq.(\ref
{eq.7})
\begin{equation}
{\cal G}(x,x^{\prime })=G(x,x^{\prime })-\kappa _{+}\left( G(0,x){\cal G}%
(0,x^{\prime })+G(x,L){\cal G}(x^{\prime },L)\right) ,
\end{equation}
where $\kappa _{\pm }=\kappa \pm ik$. It follows that
\begin{equation}
{\cal G}(0,L)=\frac{G(0,L)}{f(0)f(L)-\kappa _{+}^2G^2(0,L)}\;,\label{eq.8}
\end{equation}
where
\begin{equation}
f(z)=1+\kappa _{+}G(z,z).
\end{equation}

To proceed let us present Eq.(\ref{eq.6}) in the integral form:
\begin{equation}
G(x,x^{\prime})=G_0(x,x^{\prime})+
\int_0^Ldx_1G_0(x,x_1)v(x_1)G(x_1,x^{\prime}).\label{eq.9}
\end{equation}
Here we have introduced the unperturbed ''subbarrier'' Green function
\begin{equation}
G_0(x,x^{\prime})=-\frac 1{2\kappa }\exp \left( -\kappa \left| x-x^{\prime
}\right| \right) .\label{eq.10}
\end{equation}

The idea of further calculations can be illustrated by consideration of the
first term $G^{(1)}(x,x^{\prime })$ in the diagrammatic expansion of the
average subbarrier Green function $\left\langle G(x,x^{\prime
})\right\rangle $:
\begin{equation}
G^{(1)}(x,x^{\prime })=W\int_0^Ldx_1G_0(x,x_1)G_0(x_1,x^{\prime }).
\label{eq.t}
\end{equation}
Here the correlation function of the random part of potential $v(x)$ is
taken to be
\begin{equation}
\left\langle v(x)v(x^{\prime })\right\rangle =W\delta (x-x^{\prime })
\end{equation}
Straightforward calculation shows that
\begin{equation}
G^{(1)}(x,x^{\prime })=G_0(x,x^{\prime })\left( \frac{x^{\prime }-x}l+\frac
\gamma {2\kappa l}\right) ,\label{eq.12}
\end{equation}
where
\[
\gamma (x,x^{\prime })=2-e^{-2\kappa x}-e^{-2\kappa (L-x^{\prime })},
\]
and we have introduced the subbarrier scattering length
\begin{equation}
l=4\kappa ^2/W.
\end{equation}
The first term in brackets in Eq.(\ref{eq.12}) results from the integration
over the region $x\leq x_1\leq x^{\prime }$, where $G_0(x,x_1)G_0(x_1,x^{%
\prime })$ does not depend on $x_1$, while the second one comes from $0\leq
x_1\leq x$ and $x^{\prime }\leq x_1\leq L$, where $G_0(x,x_1)G_0(x_1,x^{%
\prime })$ decreases exponentially as a function of $x_1$. In the case that $%
x-x^{\prime }\simeq L$ the contribution from these exponential tails is
small in parameter $(\kappa L)^{-1}$. Therefore, if $\kappa L\gg 1$ the
region $x\leq x_1\leq x^{\prime }$ produces the most divergent with $L$
contribution, and integration region in Eq. (\ref{eq.t}) can be
restricted by ($x,x^{\prime }$). This simple consideration being extended
for arbitrary order in perturbation parameter $W$ leads to conclusion that
only ''$x$-ordered'' diagrams should be taken into account. This means that
we have to allow for diagrams in which the scattering points $x_i$
(variables of integration) are ordered along a straight line (Green function
line) with respect to each other and to the fixed points $x=0$ and $%
x^{\prime }=L$. For m-th order of the average Green function an ordered
diagram corresponds to the integration over the regions
\begin{equation}
0\leq x_1\leq x_2\leq \cdots \leq x_m\leq L\label{Eq.12a}
\end{equation}
if scattering points $x_i$ are numerated subsequently from the left to the
right (see Fig.1a). It is easy to see that above arguments being extended
for arbitrary order of arbitrary moments of Green function bring us to the
following selection rules: i) only diagrams containing no crossings of
scattering lines are important; ii) scattering points belonging to the same
Green function line must be ordered with respect to each other.

Selection rules being applied to the calculation of the average subbarrier
Green function give us \widetext
\begin{eqnarray}
\langle G(x,x^{\prime })\rangle &=&-\frac 1{2\kappa }\exp \left( -\kappa
\left| x-x^{\prime }\right| \right) \sum\limits_{m=1}^\infty \frac W{4\kappa
^2}\int\limits_{x_1\leq x_2...\leq x_m}dx_1...dx_m= \\
\ &=&G_0(x,x^{\prime })\sum\limits_{m=1}^\infty \frac W{4\kappa ^2}\frac{L^m%
}{m!}=G_0(x,x^{\prime })\exp \left( \left| x-x^{\prime }\right| /l\right)
\nonumber
\end{eqnarray}

\narrowtext
Function $\langle G(x,x^{\prime })\rangle $ contains factor describing an
exponential enhancement in comparison with $G_0$. It can be shown that this
is also true for all higher statistical moments $\left\langle
G^n(x,x^{\prime })\right\rangle $ as well. In the case $\kappa l\gg 1$, $%
\kappa L\gg 1$ arguments similar to above lead us to the necessity to
summarize only ordered diagrams of the type shown at Fig.1b. To perform this
calculation, let us consider a typical pair of diagrams different from each
other by permutation of two scattering lines (Fig. 2). The sum of these
diagrams can be written as
\[
...\int\limits_{x_i}^{x_j}dy\int\limits_{x_i}^ydz...+...\int%
\limits_{x_i}^{x_j}dy\int\limits_y^{x_j}dz...=...\int\limits_{x_i}^{x_j}dy%
\int\limits_{x_i}^{x_j}dz...
\]
where we have used independence of Green functions on scattering points for
ordered diagrams. As one can see, the limits of integration over $z$ do not
depend on $y$. Repeating this procedure we can add other diagrams different
by the location of one selected scattering line. This trick can be continued
till we meet scattering line of the same kind (i.e., it has scattering
points lying on the same Green function lines as that we watch for). As the
result the integrand does not depend on coordinates, and we should no longer
care of the ordering between different scattering lines. Therefore, the
problem is reduced to combinatorial governed by the fact that integration
should be restricted by the scattering points of the nearest scattering
lines. Upon applying this procedure to all possible pairs of this kind in
all perturbation orders we deduce that for $x,x^{\prime }\in (0,L)$
\begin{equation}
\frac{\left\langle G^n(x,x^{\prime })\right\rangle }{\langle G(x,x^{\prime
})\rangle ^n}=\left[ \frac{\left\langle G^2(x,x^{\prime })\right\rangle }{%
\langle G(x,x^{\prime })\rangle ^2}\right] ^{C_n^2}\;.\label{eq.16}
\end{equation}
where $C_n^2=\frac 12n(n-1)$ is the number of possibilities to combine $n$
Green functions into pairs. Calculation of $\left\langle G^2(x,x^{\prime
})\right\rangle $ is straightforward and yields
\[
\left\langle G^2(x,x^{\prime })\right\rangle =\langle G(x,x^{\prime
})\rangle ^2\exp \left( \left| x-x^{\prime }\right| /l\right) .
\]
It is convenient to define the quantity $\xi $ through the equation
\begin{equation}
\exp \left( \xi t/2\right) =G(0,L)/G_0(0,L)\label{eq.17}
\end{equation}
where the dimensionless length $t=L/l$. From Eq.(\ref{eq.16}) follows
\[
\left\langle e^{n\xi t/2}\right\rangle =\exp \left( \frac 12n(n+1)t\right) .
\]
Statistical moments of this form correspond to the Gaussian distribution low
for the random quantity $\xi (t)$:
\begin{equation}
P_t(\xi )=\sqrt{\frac t{8\pi }}\exp \left[ -\frac 18t\,(\xi -1)^2\right] .
\label{eq.19}
\end{equation}
With the same accuracy as in derivation of Eq.(\ref{eq.16}) one can find
that
\[
\left\langle G^n(0,L)G^m(x,x)\right\rangle \approx \left\langle
G^n(0,L)\right\rangle \!G_0^m(x,x).
\]
This means that $G(0,0)$ and $G(L,L)$ can be replaced by their unperturbed
values which are equal to $(-1/2\kappa )$ with the accuracy $1/\kappa l$.
After this procedure we obtain
\begin{equation}
T(\xi )=\frac{1-\cos 4\theta }{\cosh (\eta -\xi )t-\cos 4\theta }\;,
\label{eq.20}
\end{equation}
where $\eta =2\kappa l\gg 1$, $\theta =\tan {}^{-1}(\kappa /k)$. One can see
that $T(\xi =0)$ is exactly the transmission coefficient $T_0(t)$ of the
ideal barrier of dimensionless length $t$ and height $\eta $.
\begin{equation}
T_0(t)=\frac{1-\cos 4\theta }{\cosh \eta t-\cos 4\theta }
\end{equation}
Eqs.(\ref{eq.19}) and (\ref{eq.20}) bring us to the conclusion that our
problem is equivalent to the problem of transmittance of rectangular barrier
with random height where deviation $\xi $ from unperturbed value $\eta $ is
described by the Gaussian distribution function (\ref{eq.19}).

 From Eqs.(\ref{eq.19}),(\ref{eq.20}) follows that in the limit $t=L/l\gg\ 1$
\begin{equation}
\left\langle R(L)\right\rangle =\langle \frac 1{T(L)}\rangle -1=\frac{e^{L/l}%
}{T_0(L)}\;,
\end{equation}
which coincides with the result of Ref.\ \cite{Resistance}.

Eqs.(\ref{eq.19}),(\ref{eq.20}) allow us to calculate all statistical
moments of $T$
\begin{equation}
\left\langle T\,^n\right\rangle _{}=\int_{-\infty }^\infty d\xi P_t(\xi
)T^n(\xi ).\label{eq.21}
\end{equation}
For the long distance limit ($t\gg 1$) we obtain
\begin{equation}
\left\langle T\,^n\right\rangle =T_0^{\,n}(t)e^{2n\left( n+\frac 12\right)
t}\;,\;n\ll \eta /4,\;t\gg 1.\label{eq.21a}
\end{equation}
The main contribution to the large statistical moments $n\gg \eta /4$ comes
from $\xi $ of the order of $\eta $, corresponding to the strong
fluctuations of random potential when they become comparable with height of
unperturbed barrier. In this case our approach is not applicable and methods
of Refs.\ \cite{resreal} should be used. Neglecting these rear downshooting
we are able to retrieve the distribution function

\begin{equation}
P_t(\alpha )\approx \frac{\alpha ^{-\frac 34}}{\sqrt{8\pi t}}\exp \left[ -%
\frac t8\left( 1+\left( \frac{\ln \alpha }t\right) ^2\right) \right] ,\quad
\eta t\gg \ln \alpha .\label{eq.22}
\end{equation}
for normalized transmittance $\alpha (t)=T/T_0\quad (0<\alpha <1/T_0)$ that
characterizes the enhancement of the transmission rate in comparison with
that of the unperturbed barrier.

In Ref.\ \cite{enemy} the subbarrier transmission problem was attacked with
the aid of the invariant imbedding method. Due to an unestimated
approximation which enabled to solve the invariant imbedding equations there
was obtained an expression like Eq. (\ref{eq.19}) of the present paper but
different by term $-1$ in the brackets in exponent (\ref{eq.19}). This
difference becomes crucial for large $t$ and leads to wrong distribution
function of transmission coefficient.

It should be noticed that since subbarrier scattering does not randomize
phase (this follows from the fact that all functions in Eq.(\ref{eq.9}) are
real), our method is applicable also to calculation of the statistical
moments of the amplitude transmission coefficient $\tau =2k{\cal G}(0,L)$.
With the use of Eq.(\ref{eq.19}) we obtain:
\begin{equation}
\left\langle \tau ^n\right\rangle \approx \tau _0^n(t)e^{\frac 12n(n+1)t}\;,%
\text{for }n\ll \eta /2,2\kappa L\gg 1
\end{equation}
where $\tau _0$ is unperturbed value. It is readily seen that even mean
field characteristic $\left\langle \tau \right\rangle $ contains exponential
enhancement factor. This fact can be easily understood by straightforward
consideration of the Dyson equation for the true average Green function $%
\left\langle {\cal G}(x,x^{\prime })\right\rangle $:
\begin{equation}
\langle {\cal G}(x,x^{\prime })\rangle ={\cal G}_0(x,x^{\prime
})+\int_0^Ldx_1{\cal G}_0(x,x_1)\Sigma (x_1)\langle {\cal G}(x_1,x^{\prime
})\rangle \;,\label{eq.13}
\end{equation}
which can be rewritten in differential form:
\begin{equation}
\left( \frac{d^2}{dx^2}-\kappa ^2-\Sigma (x)\right) \langle {\cal G}%
(x_1,x^{\prime })\rangle =\delta (x-x^{\prime })\;.\label{eq.14}
\end{equation}
For weak disorder we can use the simplest approximation for the self-energy $%
\Sigma (x)\approx WG_0(x,x)$ which on a distance $1/\kappa \ll L$ becomes
equal to negative constant $-W/2\kappa $. Comparing Eq.(\ref{eq.14}) with
Eqs. (\ref{eq.11}), (\ref{eq.11a}) one can see that disorder causes an
effective lowering of the barrier height equal to its mean square deviation $%
-W/2\kappa $. For weak scattering ($\kappa ^2\gg \Sigma $ that is equivalent
to $\kappa l\gg 1$) the solution of Eq. (\ref{eq.13}) or (\ref{eq.14}) turns
out to be exactly the same as that obtained above by summarizing of
ordered diagrams.

To conclude, diagrammatic approach for calculation of any moments of
subbarrier Green function has been developed. The distribution function of
transmission coefficient through 1D disordered potential barrier has been
found. It is shown that an ensemble of disordered barriers is equivalent to
ensemble of rectangular barriers with random Gaussian distributed heights.
Disorder causes enhancement on average of both tunneling conductance and
resistance.


\begin{references}
\bibitem{classics}\label{sl}V. L. Berezinskii, Sov. Phys. JETP {\bf 38},
620 (1974); N. Kumar, Phys. Rev.B {\bf 31}, 5513 (1985).

\bibitem{Resistance}S. B. Haley and P. Erd\"os, Phys. Rev. B {\bf 45,}
8572 (1992).

\bibitem{resreal}I. M. Lifshitz and V. Ya. Kirpichenkov, Sov. Phys. JETP
{\bf 50}, 499 (1979); M. Raikh in {\it Mesoscopic Phenomena in Solids},
edited by B. L. Altshuler, P. A. Lee, and R. A. Webb (North-Holland,
Amsterdam, 1991), P.315.

\bibitem{enemy}J. Heinrichs, Phys. Rev. B {\bf 33}, 5261 (1986); Phys.
Rev. B {\bf 36}, 2867 (1987).
\end{references}
\end{document}